\documentclass[%
 reprint,
superscriptaddress,
 amsmath,amssymb,
 aps,
]{revtex4-2}
\usepackage{float}
\usepackage{graphicx}
\usepackage{dcolumn}
\usepackage{bm}

\usepackage{physics,amsmath,amssymb,mathtools}
\usepackage{bbold}
\DeclareMathOperator{\one}{\mathbb{1}}
\DeclareMathOperator{\zero}{\mathbb{0}}
\DeclareMathOperator{\sign}{sgn}

\newcommand{\teta}[0]{\tilde{\eta}}

\newcommand{\tsigma}[0]{\tilde{\sigma}}
\newcommand{\tlambda}[0]{\tilde{\lambda}}

\newcommand{\csp}{\alpha^{-1}}
\newcommand{\cspq}{\alpha^{-2}}

\newcommand{\redexpec}[3]{\left\langle #1 \middle| \middle| #2 \middle| \middle| #3 \right\rangle}
\begin{document}

\preprint{APS/123-QED}
\title{Application of the Variational R-matrix Method for the Dirac Equation to the Be Atom}

\author{Miguel A. Alarcón}
\email{malarco@purdue.edu}
\author{Chris H. Greene}%
\email{chgreene@purdue.edu}
\affiliation{Department of Physics and Astronomy,~Purdue University,~West Lafayette,~Indiana,~47907,~USA}%
\affiliation{Purdue Quantum Science and Engineering Institute,~Purdue University,~West Lafayette,~Indiana 47907,~USA}%

\date{\today}

\begin{abstract}
This paper presents an implementation of the non-iterative eigenchannel R-matrix method for the Dirac equation. It includes a brief introduction, implementation details, and results for the photoionization cross-section of the beryllium atom. Beryllium is a convenient test due to small but significant relativistic effects. The current calculation aligns with other R-matrix calculations and experiments. It observes the change in the Fano line shape of the $(2pnd){}^1P$ series and, reveals a previously unnoticed coupling between triplet and singlet series observable in the ground state photoionization cross-section.
\end{abstract}

\maketitle


\section{\label{sec:Intro}Introduction}
The R-matrix method is a powerful tool to study collision physics in atomic and molecular systems. It has been used to study different problems, from nuclear resonances to photoionization of atoms and molecules. The foundation of the method lies in the ability to divide configuration space into two regions: an inner region, where the interactions between the target and the projectile are strong, and an outer region, where the interactions are weak. The wave function is then expanded in terms of a set of basis functions in each region, and the coefficients of the expansion are determined by matching the wave function and its derivative at the boundary between the two regions.

Through the years, R-matrix methods have been modified and adapted to treat a wide range of problems, including time-dependent phenomena and strong laser physics~\cite{Torlina2012ARM, Guan2008TimeDept}, calculation of frequency-dependent atomic polarizability~\cite{Allison1972rmatrix} and multichannel multiple scattering for x-ray absorption spectra~\cite{kruger2018multichannel}. In this article, we will use it to describe electron-ion interactions with a focus on the description of photoionization.\\

Even in this specific use case, different formulations of the R-matrix method have been developed~\cite{burke2011r}. The eigenchannel R-matrix method~\cite{Aymar1996} has been established as a convenient implementation as, unlike the original Wigner-Eisenbud version~\cite{wigner1947higher,lane1958r}, it requires no Buttle correction~\cite{buttle1967solution} to achieve convergence with a computationally convenient sized basis. As originally formulated by Fano and Lee~\cite{FanoLee1973, Lee1974} this treatment required an iterative procedure to find solutions to the Schrödinger equation at arbitrary energies. The introduction of the direct eigenchannel variational method~\cite{Greene1983, RaseevLeRouzo, GreeneKim1988, Robicheaux1991, Aymar1996} lifts this inconvenience by expressing the solution in terms of known functions and recasting the problem in terms of algebraic equations. In the non-iterative formulation, the solution for a given energy is found by directly applying the variational principle to an expression for the logarithmic derivative at the boundary, leading to a linear generalized eigenproblem. For the sake of completeness, we have to mention that the nuclear community devised other efficient methods that do not require the Buttle correction~\cite{Lane1966, Lane1969} and that in the atomic and molecular community, there exist alternatives that do not use the variational principle~\cite{Descouvemont_2010}. For example, Baye and coauthors promote a treatment using Lagrange-Legendre functions~\cite{BayeReview2015PR}.\\

Relativistic effects have been extensively studied in R-matrix theory. Relativistic corrections have been included in the Schrödiger theory, as with the inclusion of the spin-orbit term first done in the eigenchannel formulation by Greene and Aymar~\cite{greene1991spin}, and by the use of the Breit-Pauli equation pioneered by Scott and Burke~\cite{scott1980electron}. A fully relativistic formulation had existed for almost as long as the original Wigner-Eisenbud theory thanks to Goertzel~\cite{goertzel1948resonance} and was used to study atomic systems by Chang~\cite{chang1975r,chang1977electron,chang1977relativistic}. This formulation of the method is available and widely used, in the popular DARC codes developed by Grant and collaborators~\cite{grant2007RelAMO}. 

The extension of the eigenchannel method to the Dirac equation was first explored by Hamacher and Hinze~\cite{Hamacher1991}, and was further refined and generalized by Szmytkowski~\cite{Szmytkowski1998}. However, to the authors' knowledge, no implementation of the relativistic version of the eigenchannel method has been made available. The current work presents an implementation of the method based on B-Splines. It showcases the results for the Be atom, a well-known test case for the non-relativistic R-matrix method~\cite{Omahony1985, Kim2000}, that has the advantage of having small relativistic effects that allow for the verification of the accuracy of the calculation.\\

In section II we present a summary of the theoretical background of the eigenchannel R-matrix method for the Dirac equation. In section III we present the results for the Be atom and compare them to the non-relativistic results and experimental data. Finally, in section IV we present our conclusions and future work. Two appendices containing the basics of relativistic multichannel quantum defect theory in the context of the Dirac equation are also provided. \\

The use of atomic units, where $\hbar = m_e = e = 1$, is assumed unless otherwise stated. In these units, the speed of light is $c=1/\alpha$, where $\alpha$ is the fine structure constant. 

\section{\label{sec:det_calc}Theoretical Background}
In this section we will present the fundamentals of the eigenchannel R-matrix method specialized to the two-electron case for the description of photoionization. We start with the description of the two-electron basis set, continue with the formal basics of the R-matrix method in the eigenchannel formulation, the details of the streamlined method and conclude with the formulation of the observables needed to compare with experiments. 
\subsection{Two-electron basis}
The relativistic Hamiltonian for the two valence electrons outside of the closed shell core is given by
\begin{equation}\label{eq:dir2e}
  H = h(r_1)+h(r_2) + \frac{1}{r_{12}}
\end{equation}
where $h(r_i)$ is the Dirac Hamiltonian for $i$-th electron, $r_{12}$ is the distance between the two electrons. In the current application, we exclude the Breit operator.  The explicit form of the Hamiltonian is given by

\begin{equation}\label{eq:diraceq}
  h(r_i) = \left[ -i c \bm{\alpha}\cdot \nabla_i + \beta c^2 + V(r_i) \right],
\end{equation}
where the vector $\bm{\alpha}$, represents the four-dimensional Dirac matrices 
\begin{equation}
  \bm{\alpha}_i = \begin{pmatrix}
    \zero & \sigma_i \\
    \sigma_i & \zero
  \end{pmatrix}
\end{equation} 
and the matrix $\beta$ is given by
\begin{equation}
  \beta = \begin{pmatrix}
    \one & \zero \\
    \zero & -\one
  \end{pmatrix},
\end{equation}
where the two-by-two matrices $\zero$ and $\one$ are the matrix with all zeros and the identity, respectively. The matrices $\sigma_i$ are the two-by-two Pauli matrices. 

The potential $V(r_i)$ is an effective potential energy that models the interaction of the outer electron with the doubly ionized atom
\begin{equation}\label{eq:effpot}
  V(r) = -\frac{1}{r} \left[2+(Z-2)e^{-a_1 r}+r a_2e^{-a_3 r}\right]
\end{equation}
where $Z$ is the nuclear charge and $a_i$ are parameters fitted to the single ionized atom levels. This form of the potential energy, often including a polarizability term, has been used for non-relativistic calculations \cite{Aymar1996}.

The solutions to the single electron equation 
\begin{equation}
  h(r_i) \psi_{n,\kappa}(\vec{r}_i) = E_{n,\kappa} \psi_{n,\kappa}(\vec{r}_i),
\end{equation}
in the context of atomic physics have been extensively discussed \cite{bethe2013quantum,darwin1928wave,grant2007RelAMO} and are known to be four-dimensional spinors with the form
\begin{equation}\label{eq:oneelecfunc}
  \psi_{n,\kappa,m}(\vec{r}) = \frac{1}{r} \begin{pmatrix}
    g_{n,\kappa}(r) \xi_{U,\kappa,m} (\theta,\phi) \\[2mm]
    i f_{n,\kappa}(r) \xi_{L,\kappa,m}(\theta, \phi)
  \end{pmatrix},
\end{equation}
indexed by the principal quantum number $n$ and the angular momentum numbers $\kappa$ and $m$.

The two radial functions $g_{n,\kappa}(r)$ and $f_{n,\kappa}(r)$ are solutions to the coupled set of differential equations 
\begin{equation}\label{eq:diracradsys}
  \begin{aligned}
      c  \left[ \frac{d}{dr} - \frac{\kappa}{r} \right]f_{n,\kappa} -(V(r) + \cspq -E_{n,\kappa}) g_{n,\kappa} = 0 \\[2mm]
      c \left[ \frac{d}{dr} +\frac{\kappa}{r} \right] g_{n,\kappa} +(V(r)- \cspq -E_{n,\kappa}) f_{n,\kappa} = 0
  \end{aligned}
\end{equation}
and are commonly known as the large and small components, respectively; but we will refer to them as upper and lower.\\

The functions $\xi_{U,\kappa,m}(\theta,\phi)$ and $\xi_{L,\kappa,m}(\theta,\phi)$ are the two-component angular functions for the upper and lower components, respectively. The angular momentum number $\kappa$ is related to the quantum numbers $l$ and $j$ by $\kappa = -(l+1)$ if $j=l+1/2$ and $\kappa=l$ if $j=l-1/2$. Whence, the functions in terms of coupled angular momentum states are
\begin{equation}
  \begin{aligned}
    \xi_{U,\kappa,m}(\theta,\phi) &= \ket{\left[\frac{1}{2},l\right]j,m} \\[2mm]
    \xi_{L,\kappa,m}(\theta,\phi) &= \ket{\left[\frac{1}{2},l+\sign(j-l)\right]j,m}
  \end{aligned}
\end{equation} 

For the sake of completeness, recall that the $\kappa$ quantum number is the eigenvalue of the four component operator $K= \beta\cdot(\bm{\Sigma}\cdot \bm{L} + \one_4)$, where $\Sigma_i$ are the block diagonal four-component matrices with the Pauli $\sigma_i$ in the diagonal, and $L_i$ likewise is a four component matrix with the $i$-th component of the orbital angular momentum operator in the diagonal. \\

To describe the neutral atom wave functions we use a basis of $jj$-coupled two-electron functions \cite{Aymar1996} of the form
\begin{widetext}
  \begin{equation}\label{eq:twoelecfunc}
    \begin{aligned}
      \Psi_{m_{a,b}}^{J,M} (\vec{r}_1,\vec{r}_2) =&\left[2(1+\delta_{n_a,n_b}\delta_{\kappa_a,\kappa_b})\right]^{-\frac{1}{2}}\frac{1}{r_1 r_2}\left[
        \begin{pmatrix}
        g_{n_a}(r_1) g_{n_b}(r_2) \ket{[\xi_{U,\kappa_a}(\hat{r}_1),\xi_{U,\kappa_b}(\hat{r}_2)]J,M} \\[2mm]
        i f_{n_a}(r_1) i f_{n_b}(r_2) \ket{[\xi_{L,\kappa_a}(\hat{r}_1),\xi_{L,\kappa_b}(\hat{r}_2)]J,M} 
        \end{pmatrix} \right.\\[3mm]
         &\hspace{3.5cm}-(-1)^{j_a+j_b-J} \left. \begin{pmatrix}
        g_{n_b}(r_1) g_{n_a}(r_2)  \ket{[\xi_{U,\kappa_b}(\hat{r}_1),\xi_{U,\kappa_a}(\hat{r}_2)]J,M} \\[2mm]
         i f_{n_b}(r_1) i f_{n_a}(r_2)  \ket{[\xi_{L,\kappa_b}(\hat{r}_1),\xi_{L,\kappa_a}(\hat{r}_2)]J,M} 
        \end{pmatrix}\right] 
    \end{aligned},
  \end{equation}
\end{widetext}

where the angular functions have been coupled to form the total angular momentum $J$ and its projection $M$. 

The one-electron functions were calculated using a B-Spline basis in the radial coordinate solving the coupled equations either via direct diagonalization in a finite sphere, with vanishing boundary conditions, or for specific energies by using the variational method for the ratio of the upper and lower components (see Eq.~\ref{eq:var} below). These calculations assumed that the ionic nucleus is fixed in place with infinite mass, thus the atomic units were based on the bare electronic mass. Including the effects of the finite mass of the nucleus is somewhat more involved than in the Schrödinger theory \cite{Salpeter1951ARelativistic, Salpeter1952Mass,marsch2005relativistic,marsch2006relativistic} given the necessity of coupling the spin degrees of freedom, but could be required for higher precision Rydberg physics calculations.

\subsection{Eigenchannel R-matrix method}
The fundamental derivations of the variational eigenchannel R-matrix method that we use in the current study were made by Hamacher and Hinze \cite{Hamacher1991}. For completeness, we will give a brief overview of the main formulas.

As with the non-relativistic method, the starting point is a variational form of the quantity that defines a solution of the equation in a finite volume. For the Schrödinger equation, this quantity is the logarithmic derivative of the scaled wave function at the sphere boundary \cite{Greene1983}. For the Dirac equation, the quantity is the ratio of the lower and upper components of the wave function, evaluated at the boundary radius. This can be derived by evaluating the asymmetry of the one-electron Hamiltonian in the finite volume for two solutions and determining the condition needed for it to vanish. \\

Let $b$ be the ratio of the lower to the upper component at the boundary of a spherical box of radius $R$ for the solution at energy $E$. Defining the Bloch operator $\Delta$ and boundary operator $T$ as
\begin{equation}
\begin{aligned}
      \Delta = i \csp \begin{pmatrix}
    \zero & \sigma_r \\
    \zero & \zero\\
  \end{pmatrix} \delta(r-R), \\[2mm]
  T = \csp \begin{pmatrix}
    \one & \zero \\
    \zero & \zero
\end{pmatrix} \delta(r-R),
\end{aligned}
\end{equation}
the exact expression for $b$ given a solution $\psi$ is
\begin{equation}\label{eq:var}
  b[\psi] = \frac{\bra{\psi}h+\Delta-E\ket{\psi}}{\bra{\psi}T\ket{\psi}}.
\end{equation}

One can show that this expression, viewed as a functional of a trial solution, is variational, i.e. constant to first order in variations to the trial solution.\cite{Hamacher1991} If the trial solution at energy $E$ is expanded in terms of some basis as $\ket{\psi}=\sum_k c_k \ket{y_k}$, the coefficients must solve the generalized eigenvalue problem:
\begin{equation}\label{eq:genEig}
  \sum_j (h+\Delta-E)_{i,j} c_j = b \sum_j T_{i,j} c_j
\end{equation}
where the indices $i,j$ run over the basis functions. There will be a value of $b$ and a set of coefficients for each possible solution. In this case, having a single particle with a single channel there is a single solution for a given energy. In the case of multiple channels, the number of obtained solutions will be given by the number of open and weakly closed channels.\\

In the multielectron case, the only modifications needed in the equations are changing the Bloch and boundary operator to a sum of single electron operators and changing the Hamiltonian accordingly. Here we specialize in the two-electron case and use the Hamiltonian in Eq.~\ref{eq:dir2e}. The literature commonly defines the matrices $\Gamma$, representing Hamiltonian plus the Bloch operator, Eq.~\ref{eq:genEig}, and $\Lambda$ representing the surface operator.

\subsection{The Streamlined Method}
To describe photoionization one needs to use a basis that allows one electron to exit the finite volume of the R-matrix box. A basis composed of two-electron functions, with different boundary conditions, has been used when solving the Schrödinger equation \cite{Greene1983, Aymar1996}. We use the functions shown in Eq.~\ref{eq:twoelecfunc} and define closed configurations -in which the upper component vanishes for both electrons- and open configurations -in which the upper component does not vanish at the sphere boundary for only one of the electrons.

As was shown in \cite{Greene1983, Aymar1996}, and we summarize here, a convenient and efficient way in which to solve the generalized eigenvalue problem from Eq.~\ref{eq:genEig} is to split the coefficients into an open and a closed subspace. Let $N_o$ and $N_c$ denote the number of open and closed type configurations, respectively, and let $\left\{\ket{y^o_k}\right\}_{k=1}^{N_o}$ and $\left\{\ket{y^c_k}\right\}_{k=1}^{N_c}$ be the two-electron basis functions for the open and closed type configurations, respectively. 

The solution wave function corresponding to the eigenvalue $b_{\zeta}$ can be written in terms of two vectors of coefficients $c^o$ and $c^c$ as
\begin{equation}\label{eq:multichanbeta}
  \ket{\Psi_\zeta} = \sum_{k=1}^{N_o} c^o_k \ket{y^o_k} + \sum_{k=1}^{N_c} c^c_k \ket{y^c_k},
\end{equation}

Since the upper component of closed-type configurations vanishes at the boundary, the matrix elements of the boundary operators vanish and the generalized eigenvalue problem can be written as 

\begin{equation}
  \begin{aligned}
    \Gamma_{oc} c^c - E O_{oc} c^c + \Gamma_{oo} c^o - E O_{oo} c^o   &= b_\zeta \Lambda_{oo} c^o \\[2mm]
    \Gamma_{cc} c^c - E O_{cc} c^c + \Gamma_{co} c^o  - E O_{co} c^o &= 0 \\
  \end{aligned}
\end{equation}
where the matrix $O$ is the overlap of the configurations. In general, the basis is constructed such that closed-type configurations are orthogonal, but the open-type are not orthogonal either between them or with the closed type.

From the second equation, one can solve for the closed-type coefficients in terms of the open-type coefficients to yield 
\begin{widetext}
  \begin{equation}
      \left[ \Gamma_{oo} - E O_{oo} - (\Gamma_{oc}-E O_{oc}) \left( \Gamma_{cc} - E O_{cc} \right)^{-1} (\Gamma_{co}-EO_{co}) \right] c^o\\
       = b_\zeta \Lambda_{oo} c^o,
  \end{equation}
\end{widetext}
which is the equation we solve numerically. 

This recasting makes the calculation more efficient, since in general the number of closed-type configurations needed for a calculation to converge is much larger than the number of open-type configurations; this step thus reduces the size of the problem. One final simplification can be achieved by noticing that since the closed configurations are orthogonal the overlap matrix $O_{cc}$ is the identity, and the matrix $\Gamma_{cc}$ can be diagonalized making the inversion a trivial step.

\subsection{Physical Solutions, Dipole moments and Cross Sections}

To extract the physical solutions in all space, the solution found for the finite volume has to be matched with the solution in the outer region. In the case of a neutral atom, as one of the electrons reaches the box boundary the interaction with the residual core is that of a Coulomb potential. Therefore, when the position of one of the electrons is set at the box boundary the function is expanded in terms of Coulomb functions.

In the Dirac equation, these solutions have been extensively studied \cite{bethe2013quantum,grant2007RelAMO} and a generalized quantum defect theory was developed by Chang \cite{chang1993general}. The functions to define the quantum defect used in the current implementation are specified in the appendix. There, we also show the basics of the generalized quantum defect theory.

The channel functions, containing the angular degrees of freedom of both electrons and the radial component of the innermost electron $r_1$, are represented by the spinors
\begin{equation}
  \Phi_{j,J,M} = \frac{1}{r_1}%
  \begin{pmatrix}
   g_{n_a}(r_1) \ket{[\xi_{U,\kappa_a}(\hat{r}_1),\xi_{U,\kappa_b}(\hat{r}_2)]J,M} \\[2mm]
   i f_{n_a}(r_1) \ket{[\xi_{L,\kappa_a}(\hat{r}_1),\xi_{L,\kappa_b}(\hat{r}_2)]J,M}\\ 
  \end{pmatrix} 
\end{equation}
where $j$ is an index that summarizes all the one-electron quantum numbers. 

When the outer electron is outside the finite volume, meaning $r_2>R$, the channel amplitude in the eigenchannel solution is given by
\begin{equation}
  r_2 \bra{\Phi_{j,J,M}}\ket{\Psi_\zeta} = \frak{f}^o_j(r_2) \mathcal{I}_{j,\zeta} - \frak{g}^o_j(r_2) \mathcal{J}_{j,\zeta},
\end{equation}
where the subindex $j$ in the Coulomb functions indicate evaluation at the channel energy $\epsilon_j=E-E_{n_a}$ and with the outer electron angular quantum number $\kappa_b$. A channel is open if $\epsilon_j$ is larger than the electron rest energy, and is closed otherwise. \\

The reaction matrix is defined by $K=\mathcal{J} \mathcal{I}^{-1}$, and the eigenchannel solutions $\Psi_\rho$ are defined by
\begin{equation}
  r_2 \bra{\Phi_{j,J,M}}\ket{\Psi_\rho} \underset{r_2\to\infty}{\rightarrow} \begin{cases}
    T_{j\rho}\left(\mathfrak{f}_j \cos \pi \tau_\rho -\mathfrak{g}_j\sin \pi \tau_\rho\right) & \text{if } \epsilon_j>\cspq \\[2mm]
    0 & \text{otherwise}
  \end{cases}
\end{equation}
As detailed in the appendix, these solutions are determined by finding the correct linear combination of functions that satisfy these boundary conditions. 

To compute cross sections, we use the eigenchannel solutions to compute reduced dipole amplitudes with an initial state $\Psi_o$ in the length form
\begin{equation}
  d_\rho = \redexpec{\Psi_\rho}{\vec{r}_1+\vec{r}_2}{\Psi_o},
\end{equation}
and in the velocity form
\begin{equation}
  v_\rho = \frac{c}{E-E_o} \redexpec{\Psi_\rho}{\vec{\bm{\alpha}}_1+\vec{\bm{\alpha}}_2}{\Psi_o}.
\end{equation}
the level of disagreement between the two forms of the dipole operator represents a measure of the inaccuracy of the numerical calculation.

The physical reduced dipole matrix elements, with the states that reach the detector with an electron on each independent channel, are given by

\begin{equation}
  d_j = \sum_{\rho} T_{j\rho} d_\rho \exp\left[i(\pi \tau_\rho-\gamma_j \pi/2-\sigma_j)\right]
\end{equation}
where $\sigma_j$ is the Columb relativistic phase shift for the channel $j$, and $\gamma_j$ is a parameter related to the angular quantum number, both are defined in the appendix. In that appendix, we also define the linear combination of eigenchannel solutions that give the physical solutions, as well as the density of states in the closed channel components that we refer to as $Z$ coefficients \cite{Aymar1996} and will aid in the classification of autoionizing resonances.

The total isotropic cross-section is given by 

\begin{equation}
  \sigma = \frac{4\pi^2}{3} \frac{E-E_o}{2J_o+1} \alpha \sum_{m} \left| d_m \right|^2
\end{equation}
where the initial state $\Psi_o$ is assumed to have angular momentum $J_o$.
\section{Results for Beryllium}
\begin{table}[t]
  \caption{\label{tab:potparam}Parameters for the effective potential. These parameters were optimized in previous work\cite{Aymar1996} to reproduce the non-relativistic ionization levels using an $LS$-coupled version of the R-matrix code .
  }
  \begin{ruledtabular}
    \begin{tabular}{lccr}
      $Z$ & $a_1$ & $a_2$ & $a_3$ \\
      \hline
      4 & 6.9010 & 8.9581 & 5.0798
    \end{tabular}
  \end{ruledtabular}
  
\end{table}

We used the beryllium atom as an initial test for the method as it was one of the first study cases for the non-relativistic method \cite{Omahony1985, Kim2000}, and will have small relativistic effects making it a reliable test of the accuracy of the current implementation. We compare the Dirac results with Schrödinger results that include the spin-orbit corrections. In addition, we can also compare with other relativistic methods \cite{Hwang1991} and with available experimental data \cite{Olalde-Velasco2007, Wehlitz2003}.

The relativistic calculation used a 20-bohr R-matrix box and 20~orbitals per partial wave, divided into 18~closed and 2~open. The orbitals were computed using a B-Spline basis with 150 intervals variably distributed. A square root mesh (i.e. setting $r\equiv x^2$, equally spaced in $x$) was used for the first 84 intervals, after which they were distributed uniformly in $r$ until the end of the box. The integrals were computed using order 8~splines with a 16~point Gaussian quadrature per interval.

\begin{figure}[b]
  \includegraphics[width=0.45\textwidth]{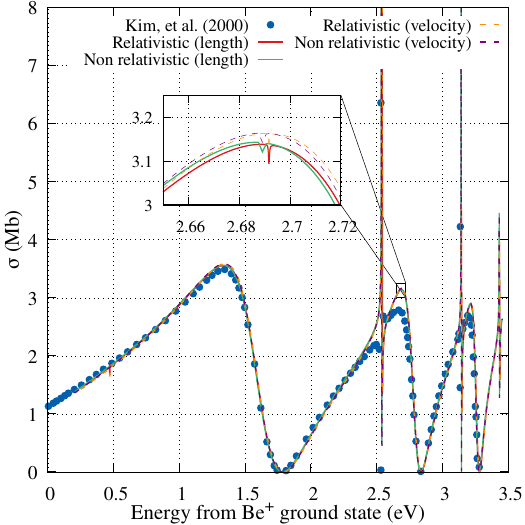}
  \caption{\label{fig:be_below2p12}%
  Photoionization cross section for the ground state of Be below the $1s^2 2p_{1/2}$ threshold. The relativistic and non-relativistic calculations are shown in the length (solid) and velocity gauges (dashed). Color version online. The dotted line is the data obtained from \cite{Kim2000}.}
\end{figure} 

The interaction of the electrons with the residual Be$^{2+}(1s^2)$ core was modeled via the potential in Eq.~\ref{eq:effpot} with the parameters shown in Tab.~\ref{tab:potparam}, which were optimized to obtain the non-relativistic ion levels \cite{Aymar1996}. The ionization thresholds are shown in Tab.~\ref{tab:levels}, where the levels obtained with the relativistic code, the non-relativistic $jj$-coupled code, and the experimental values \cite{NIST_ASD} are compared. The $jj$-coupled code, described in detail in \cite{Aymar1996}, was run with the same box size and number of orbitals as the relativistic code. 

\begin{table}[b]
  \caption{\label{tab:levels}%
  Ionization thresholds (in atomic units) relative to the double ionization threshold.}
  \begin{ruledtabular}
    \begin{tabular}{lccr}
      Level & Relativistic & Non-relativistic & Exp. \cite{NIST_ASD} \\[0.5mm]
      \hline
      $1s^2 2s_{1/2}$ & $-0.669263$ & $-0.669113$ & $-0.669248$ \\[1.5mm]
      $1s^2 2p_{1/2}$ & $-0.523692$ & $-0.523661$ & $-0.523770$ \\[1.5mm]
      $1s^2 2p_{3/2}$ & $-0.523634$ & $-0.523604$ & $-0.523740$ \\[1.5mm]
      $1s^2 3s_{1/2}$ & $-0.267667$ & $-0.267625$ & $-0.267233$ \\[1.5mm]
      $1s^2 3p_{1/2}$ & $-0.229824$ & $-0.229812$ & $-0.229582$ \\[1.5mm]
      $1s^2 3p_{3/2}$ & $-0.229806$ & $-0.229794$ & $-0.229574$ \\[1.5mm]
      $1s^2 3d_{3/2}$ & $-0.222389$ & $-0.222387$ & $-0.222478$ \\[1.5mm]
      $1s^2 3d_{5/2}$ & $-0.222387$ & $-0.222384$ & $-0.222476$ \\[1.5mm]
    \end{tabular}
  \end{ruledtabular}
\end{table}

Photoionization cross sections for ground state ionization were calculated in length and velocity gauges. Good agreement between the two gauges, and between relativistic and non-relativistic calculations is found across all energy ranges. To ensure the stability of the numerics different open and weakly closed channels are included depending on the energy region.

\begin{figure}[t]
  \includegraphics[width=0.45\textwidth]{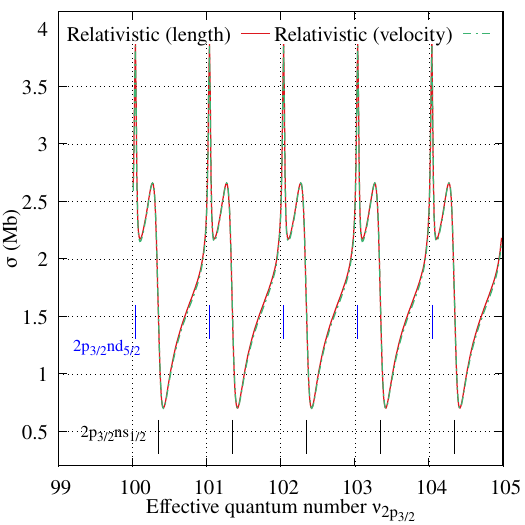}
  \caption{\label{fig:be_below2p32}%
  The photoionization cross-section of the Be ground state between the $1s^2 2p_{1/2}$ and $1s^2 2p_{3/2}$ thresholds showcases the narrow $2p_{3/2} n d_{5/2}$ and broad $2p_{3/2} n s_{1/2}$ autoionizing states. The relativistic results are shown in the length (solid) and velocity (dashed) gauges. The non-relativistic calculations agree exactly with these calculations and for clarity are not shown.}
\end{figure}

\begin{figure*}[th]
  \includegraphics[width=2\columnwidth]{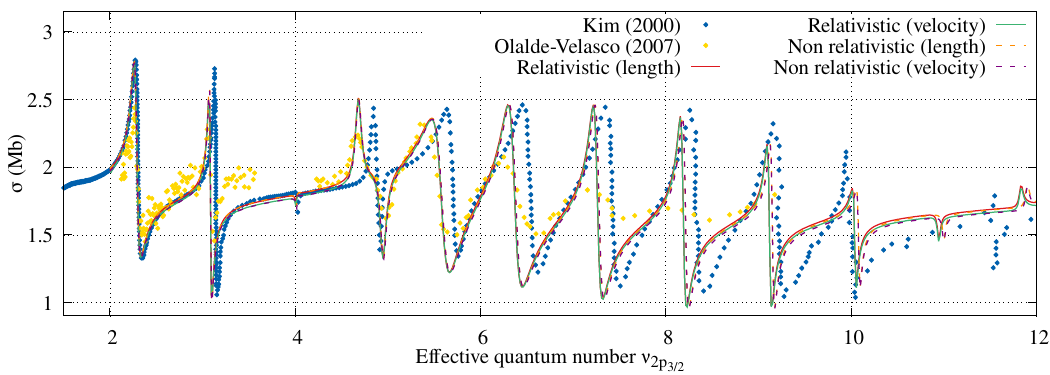}
  \caption{\label{fig:be_below3s12}%
  The Be ground state photoionization cross section at final state energies just below the $1s^2 3s_{1/2}$ threshold. The relativistic and non-relativistic calculations are shown in the length (solid) and velocity gauges (dashed). Color version online. The points are the data obtained from \cite{Kim2000}.}
\end{figure*}
For the region below the $1s^2 2p_{1/2}$ threshold, 7 channels were used. The open $2s_{1/2} \epsilon p_{1/2}$ and $\epsilon p_{3/2}$, and the closed $2p_{1/2} n s_{1/2}$ and $nd_{3/2}$, $2p_{3/2} n s_{1/2}$, $n d_{3/2}$ and $n d_{5/2}$. The results are shown in Fig.~\ref{fig:be_below2p12} and show great agreement between the two approaches, comparing favorably to the $LS$ coupled calculation \cite{Kim2000}, and is certain to be accurate given the agreement between the two gauges. Two distinguishable features appear to manifest the relativistic effects. The most glaring is the change in the Fano line shape that the $2p nd$ ionization series shows. This shape is not a new finding, it has been observed in other relativistic R-matrix studies of the atom \cite{chu2009photoionization} and could even be traced back to atomic opacity calculations \cite{tully1990atomic}. Some experimental evidence for the shape of this resonance exists \cite{Wehlitz2003}, but the level of detail needed to confirm it is not available.

The second feature, that does not seem to have been pointed out before, is a very subtle detail on the cross-section signal. Near the peak of the cross section corresponding to the $2pns ^1S_1$ series one can observe a very narrow and shallow window resonance. We confirm the existence of the resonance by analyzing the time delay at that energy and the $Z$ coefficients~\cite{Aymar1996} (see appendix, equation \ref{eq:zcoeffeq}), and noticing that a Lorentzian peak in the time delay aligns with the window. A detailed view is presented in the inset of Fig.~\ref{fig:be_below2p12}. Looking at the $Z$ coefficients allowed us to determine that this resonance is dominated by the $2p_{1/2}ns_{1/2}$ and the $2p_{3/2}ns_{1/2}$ channels, suggesting that it has to be a relativistic effect that couples the $^1S_1$ and the $^3S_1$ series. We further confirmed this by analyzing the photoionization cross-section for the $(2p^2) {}^3P_1$ state that does couple to this series.
A similar feature can be found around the location of the $(2pnd) {}^3P_1$ series, a very small feature is observable at the location of the $(2pnd) {}^3D_1$ series, which should be doubly forbidden in $LS$-coupling as there is no $^3D_1$ continuum to decay to and violates the $|\Delta L|\le 1$ and $\Delta S=0$ dipole selection rules. It was observed that the presence of the $^3D_1$ is even stronger in the photoionization of the $(2p^2) {}^3P_0$ state.\\

With the same number of channels, we explored the region between the two spin-orbit split thresholds $1s^2 2p_{1/2}$ and $1s^2 2p_{3/2}$. The results are shown in Fig.~\ref{fig:be_below2p32}. This plot serves as a correction to Fig.~4 of \cite{Aymar1996} as the authors apparently mislabeled the horizontal effective quantum number axis. Making a similar analysis of the time delays allowed us to make the labeling we show in the plot. Observe that even if the time delays show three resonances in one cycle of $\nu_{2p_{3/2}}$, only two resonances appear in the cross-section, as the dipole coupling to the ground state is very small for the $2p_{3/2} nd_{3/2}$ series. \\

For energies above the $1s^2 2p_{3/2}$ threshold, we added 7 channels, similar to the ones from before but changing the core state from $2s$ and $2p$ to $3s$ and $3p$. For energies just below the $3s_{1/2}$ threshold just including these channels proved sufficient. For slightly higher energies, one needed to include the 6 channels attached to the $3d$ thresholds: $3d_{3/2} np_{1/2}$, $np_{3/2}$ and $nf_{5/2}$, and $3d_{5/2} np_{3/2}$, $nf_{5/2}$ and $nf_{7/2}$. \\
\begin{figure}
  \includegraphics[width=0.45\textwidth]{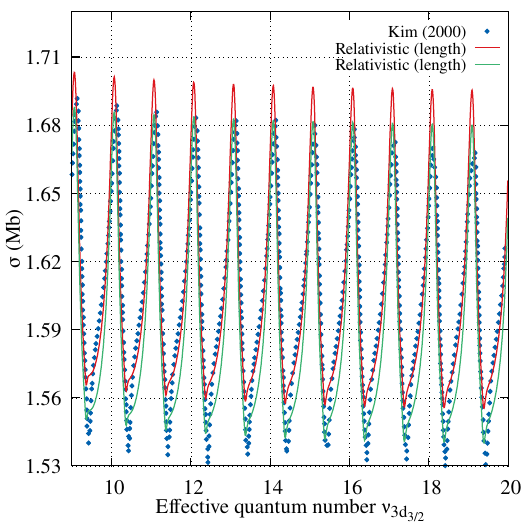}
  \caption{\label{fig:be_above3s12}%
  Photoionization cross section for the ground state of Be below the $1s^2 3d_{3/2}$ threshold. The relativistic and non-relativistic calculations are shown in length (solid) and velocity gauges (dashed), with the points denoting the $LS$ coupled calculation \cite{Kim2000}. Color version online. Again, the non-relativistic calculation agrees exactly with the relativistic one, and for clarity is not shown.}
\end{figure}

The photoionization cross section obtained just below the $3s_{1/2}$ threshold is shown in Fig.~\ref{fig:be_below3s12}. Again, the relativistic and non-relativistic calculations are in great agreement, as well as the comparison with the calculation $LS$-coupled calculation \cite{Kim2000}. The figure also includes the experimental data from \cite{Olalde-Velasco2007} which shows good agreement with the calculations. One of the issues that Olalde, et al. \cite{Olalde-Velasco2007} criticize about the R-matrix result is that the peaks near the threshold seem to be overestimated. The relativistic theory gives a better result, in both the width of the resonances and the peak position. 

For energies above the $3s_{1/2}$, where all 13 channels are included, good agreement between the relativistic and non-relativistic calculations is found even though for these energies the calculation apparently has poorer convergence, as is evident in Fig.~\ref{fig:be_above3s12} where the difference between the two gauges is more pronounced. This is probably attributable to the size of the R-matrix box being a little too small, which does not accurately contain the physical $3d$ orbitals of the ion.

\section{Conclusions}
In this study, we have shown an implementation of the relativistic non-iterative eigenchannel R-matrix method. The main ideas of the method are presented and the implementation has been tested for beryllium where multiple studies are available for comparison, both in theory and experiment.

The results of the fully relativistic calculations agree well with the experiment and with the spin-orbit corrected Schrödinger theory. Some especially sensitive features of the spectrum that change when relativistic effects are included, such as the Fano line shape of the $2pnd$ series, were observed. It was pointed out that both the $(2pnd)^3D_1$ and the $(2pns)^3P_1$ series appear in the ground state photoionization cross section thanks to relativistic effects (primarily spin-orbit coupling), even if the former should be completely decoupled from the continuum in $LS$ coupling.

The method presented will be used to analyze systems where the relativistic effects are bound to be more pronounced and where theoretically accurate predictions can lead to the realization of new experiments. Specifically the method is to be used to study radium where a complete characterization of the autoionization spectrum is yet to be done, and where new experiments \cite{fan2023laser} could determine the accuracy of the calculations.

\section{Acknowledgments}

M.A. thanks F. Robicheaux for helpful discussions in the project's early development. This research was possible thanks to support from the U. S. Department of Energy, Office of Science, Office of Basic Energy Sciences, under Award No. DE-SC0010545.

\bibliography{apssamp}
\begin{widetext}
\appendix

\section{Relativistic Quantum Defect Theory}
\subsection{Single Channel Formulation}
 In this appendix, we want to set the convention used in this work for the generalized quantum defect theory we used for the Dirac equation. The presentation follows closely that of Chang \cite{chang1993general}, although we use a slightly different form for the normalization of the coulomb functions and base our negative energy multichannel analysis on the analytic functions. We will only present the results for the Coulomb potential.

 As a starting point, define the two solutions of the coupled system of Eq.~\ref{eq:diracradsys} with the potential $V(r)=-Z/r$. Just as with the non-relativistic case, there are two solutions one that is regular at the origin and one that diverges. A generic form for these solutions is

  \begin{equation}
  \begin{aligned}
    r\Psi_1 = \frac{C}{\gamma+i \eta_1} e^{-i\lambda r} r^{\gamma}\begin{pmatrix}
      \sqrt{\epsilon+1} \left[-(\kappa+i\eta_2)\Phi(\gamma+i\eta_1,1+2\gamma,2i\lambda r) +(\gamma+i\eta_1) \Phi(1+\gamma+i\eta_1,1+2\gamma,2i\lambda r)\right] \xi_U \\
      \sqrt{\epsilon-1}\left[-(\kappa+i\eta_2)\Phi(\gamma+i\eta_1,1+2\gamma,2i\lambda r) - (\gamma+i\eta_1) \Phi(1+\gamma+i\eta_1,1+2\gamma,2i\lambda r)\right] \xi_L
   \end{pmatrix}\\[5mm]
   r\Psi_2 = \frac{D}{\gamma-i \eta_1} e^{-i\lambda r} r^{-\gamma}\begin{pmatrix}
    \sqrt{\epsilon+1} \left[(\kappa+i\eta_2)\Phi(-\gamma+i\eta_1,1-2\gamma,2i\lambda r) +(\gamma-i\eta_1) \Phi(1-\gamma+i\eta_1,1-2\gamma,2i\lambda r)\right] \xi_U \\
    \sqrt{\epsilon-1}\left[(\kappa+i\eta_2)\Phi(-\gamma+i\eta_1,1-2\gamma,2i\lambda r) -(\gamma-i\eta_1) \Phi(1-\gamma+i\eta_1,1-2\gamma,2i\lambda r)\right] \xi_L
  \end{pmatrix}
  \end{aligned}
  \end{equation}

 where $\Psi_1$ and $\Psi_2$ are a regular and an irregular solution, respectively, and $C$ and $D$ are normalization constants. The functions are given in terms of the confluent hypergeometric function of the first kind $\Phi(a,b,z)$, and the following parameters: $\gamma=\sqrt{\kappa^2-(Z\alpha)^2}$, $\epsilon=E \alpha^2$, $\lambda=\sqrt{\epsilon^2-1}/\alpha$, $\eta_1 =Z\alpha \epsilon/\sqrt{\epsilon^2-1}$ and $\eta_2= Z\alpha/\sqrt{\epsilon^2-1}$. Since we will be concerned with the radial solutions, we will drop the two-component angular solutions. We will refer to the two-component radial solution by lower case $\psi$ and the complete solution by upper case $\Psi$.

 A regular solution that is analytic in energy,$\Psi_1^o$, can be obtained by setting 
 \begin{equation}
  C =\frac{(\gamma+i\eta_1)}{[\gamma-\kappa+i(\eta_1-\eta_2)]\sqrt{1+\epsilon}} \frac{(2Z)^\gamma}{\Gamma(1+2\gamma)}\sqrt{\frac{(\kappa-\gamma)^2}{Z}}.
 \end{equation} 

 Similarly, an irregular solution that is analytic in energy, $\Psi_2^o$, can be obtained by setting
 \begin{equation}
  D =\frac{(\gamma+i\eta_1)}{[\gamma+\kappa-i(\eta_1-\eta_2)]\sqrt{1+\epsilon}} \frac{(2Z)^{-\gamma}}{\Gamma(1-2\gamma)}\sqrt{\frac{(\kappa+\gamma)^2}{Z}}.
 \end{equation} 

 For \textit{positive energy}, $\epsilon>1$, these solutions have the following asymptotics

 \begin{equation}
  \begin{aligned}
  r \psi_1^o \underset{r\to\infty}{\to} 2 \sqrt{\frac{(\gamma-\kappa)^2}{Z}}\frac{(Z/\lambda)^\gamma}{|\Gamma(\gamma+i\eta_1)|} \frac{e^{-\pi \eta_1/2}}{\left|\gamma-\kappa+i (\eta_1-\eta_2)\right|}  \begin{pmatrix}
    \cos\left(\lambda r + \eta_1 \ln 2\lambda r -\gamma \pi/2-\sigma_{\gamma,\epsilon} \right) \\
    -\sqrt{\frac{\epsilon-1}{\epsilon+1}}\sin\left(\lambda r + \eta_1 \ln 2\lambda r -\gamma \pi/2 -\sigma_{\gamma,\epsilon} \right)
\end{pmatrix} \\[3mm]
r \psi^o_2 \underset{r\to\infty}{\to} 2 \sqrt{\frac{(\gamma+\kappa)^2}{Z}}\frac{(Z/\lambda)^{-\gamma}}{|\Gamma(-\gamma+i\eta_1)|} \frac{e^{-\pi \eta_1/2}}{\left|\gamma+\kappa-i (\eta_1-\eta_2)\right|}  \begin{pmatrix}
  -\cos\left(\lambda r + \eta_1 \ln 2\lambda r +\gamma \pi/2 -\tsigma_{\gamma,\epsilon} \right)\\
   \sqrt{\frac{\epsilon-1}{\epsilon+1}}\sin\left(\lambda r + \eta_1 \ln 2\lambda r +\gamma \pi/2 -\tsigma_{\gamma,\epsilon} \right)
\end{pmatrix}
\end{aligned}
 \end{equation}
The phase parameter are defined as $\sigma_{\gamma,\epsilon}~=\arg [\Gamma(\gamma+i \eta_1)] + \arg[\gamma-\kappa+i(\eta_1-\eta_2)]$ and $\tilde{\sigma}_{\gamma,\epsilon} = \arg[\Gamma(-\gamma+i\eta_1)]+\arg[\gamma+\kappa - i(\eta_1-\eta_2)]$. Now, to appropriately describe photoionization we need the functions that are energy normalized. The correct normalization constant is found by using the technique shown in Eq.~52 in \cite{Fano1970Quantum}. The asymptotic form of the energy-normalized functions is given by
  \begin{equation}
    \begin{aligned}
    r \psi_1 \underset{r\to\infty}{\to} \begin{pmatrix}
      \sqrt{\frac{\epsilon+1}{\lambda \pi}} \cos(\lambda r + \eta_1 \ln 2\lambda r -\gamma \pi/2 -\sigma_{\gamma,\epsilon}) \\
      -\sqrt{\frac{\epsilon-1}{\lambda \pi}} \sin(\lambda r + \eta_1 \ln 2\lambda r -\gamma \pi/2 -\sigma_{\gamma,\epsilon}) \\
  \end{pmatrix} \\[3mm]
    r \psi_2 \underset{r\to\infty}{\to} \begin{pmatrix}
      -\sqrt{\frac{\epsilon+1}{\lambda \pi}} \cos(\lambda r + \eta_1 \ln 2\lambda r + \gamma \pi/2 -\tilde{\sigma}_{\gamma,\epsilon}) \\
      \sqrt{\frac{\epsilon-1}{\lambda \pi}} \sin(\lambda r + \eta_1 \ln 2\lambda r + \gamma \pi/2 -\tilde{\sigma}_{\gamma,\epsilon})\\
  \end{pmatrix}
  \end{aligned}
  \end{equation}

Using these solutions one could define a quantum defect theory but a more convenient basis on which to define it is one where the irregular solution is $\pi/2$ out of phase with the regular solution. This other irregular solution, $\psi_3$, is given in terms of the difference of the phases $\psi_3 = \cot( \sigma -\tilde{\sigma}+\pi \gamma) \psi_1 + \csc( \sigma -\tilde{\sigma}+\pi \gamma) \psi_2$, which by using the gamma inversion formula, can be expressed in a form that is easier to analyze as a function of energy
\begin{equation}
  \psi_3 = \left(\cot 2\pi\gamma -  e^{-2\pi\eta_1} \csc 2\pi\gamma \right) \psi_1 - \csc 2\pi\gamma \frac{2 \pi e^{-\pi \eta_1}}{\sqrt{\kappa^2+\eta_2^2} |\Gamma(\gamma+i\eta_1)\Gamma(-\gamma+i\eta_1)|} \psi_2
\end{equation}
which has the asymptotic form
\begin{equation}
    r\psi_3 \underset{r\to\infty}{\to} \begin{pmatrix}
        \sqrt{\frac{\epsilon+1}{\lambda \pi}}\sin(\lambda r + \eta_1 \ln 2\lambda r -\gamma \pi/2 -\sigma_{\gamma,\epsilon}) \\
        \sqrt{\frac{\epsilon-1}{\lambda \pi}}\cos(\lambda r + \eta_1 \ln 2\lambda r -\gamma \pi/2 -\sigma_{\gamma,\epsilon})
    \end{pmatrix}.
\end{equation}
Before continuing, notice that from the asymptotic form, one can see that the Wronskian of the two solutions is constant with value $W(\psi_1,\psi_3)=\alpha/\pi$. To be consistent with the rest of the literature on quantum defect theory, we will refer to these two functions as $\frak{f}:=\psi_1$ and $\frak{g}:=\psi_3$. \\

It is desirable to find a solution that is also analytic in energy and has this same Wronskian. The function we just built gives us a hint on how to do this, especially when noticing that the Wronskian of the two analytic solutions is $W(\psi^o_1, \psi^o_2)=-\alpha/\pi \csc 2\pi\gamma$. First, notice that the analytic regular solution and the energy normalized solutions are related by a scaling factor
\begin{equation}
  B(\epsilon) = \frac{\Gamma(\gamma+i\eta_1)\Gamma(\gamma-i\eta_1)}{2 \pi } \frac{\kappa \epsilon -\gamma}{\kappa -\gamma} \eta_2^{-2 \gamma+1} e^{\eta_1 \pi} \Rightarrow \psi^o_1 = B(\epsilon)^{1/2} \frak{f}
\end{equation}
 which when combined with the expression for the phase difference in terms of energy used when defining $\frak{g}$ allows us to define an intermediate function 
 \begin{equation}
  \overline{\psi}_3 = B(\epsilon)(\cot(2\pi \gamma)-\csc(2\pi\gamma) e^{-2\pi\eta_1}) \psi^o_1 - \csc(2\pi\gamma) \psi_2^o
 \end{equation}
 which has $W(\psi^o_1,\overline{\psi}_3)=\alpha/\pi$, but is not analytic due to the factor $B(\epsilon)e^{-2\pi\eta_1}$. But, similar to what is done in the non-relativistic case, the analytic function is obtained by subtracting this non-analyticity
 \begin{equation}
  \psi_3^o := \overline{\psi}_3 + B(\epsilon) \csc(2\pi\gamma) e^{-2\pi \eta_1} \psi^o_1.
 \end{equation}
 This can be done for the case of a non-fractional $Z$, where the factor $\gamma$ is not an integer. When $\gamma$ is an integer, similar problems to the ones found in the non-relativistic case arise, and the solution is outlined in \cite{chang1993general}. Again, to fall in line with the rest of the literature we will refer to these two functions as $\frak{f}^o:=\psi^o_1$ and $\frak{g}^o:=\psi^o_3$.

 Going to negative energies, $\epsilon<1$, we use the most natural branch when doing the analytical continuation of the parameters $\lambda \to i\tilde{\lambda}, \quad \eta_1 \to -i \tilde{\eta}_1, \quad \eta_2\to-i\tilde{\eta_2}$. To simplify the asymptotic forms, we define the parameters
 
 \begin{equation}
  \begin{aligned}
       A(\epsilon) &= \frac{\Gamma(\gamma+\teta_1)}{\Gamma(1+\teta_1-\gamma)} \frac{\kappa \epsilon-\gamma}{\kappa-\gamma} \teta_2^{1-2\gamma}, \\[3mm] 
       D(\epsilon) &= \frac{\Gamma(\gamma+\teta_1) \Gamma(1+\teta_1-\gamma) (\gamma-\kappa+\teta_1-\teta_2)}{\pi (\gamma-\kappa-\teta_1+\teta_2)}.
  \end{aligned}
  \end{equation}

  and find, with some algebra, that the asymptotic forms of the analytic solutions are

  \begin{equation}
  \begin{aligned}
    r\psi^o_{1} \to \frac{A^{-1/2}(\epsilon)}{\sqrt{2\pi \tlambda}}  
    \begin{pmatrix} \sqrt{\epsilon+1} \left[-e^{\tlambda r} (2\tlambda r)^{-\teta_1} D(\epsilon)^{1/2} \sin \pi (\teta_1-\gamma) + e^{-\tlambda r} (2\tlambda r)^{\teta_1} D(\epsilon)^{-1/2} (-1)^{-\gamma+\teta_1}\right] \\[3mm] 
    \sqrt{1-\epsilon} \left[-e^{\tlambda r} (2\tlambda r)^{-\teta_1} D(\epsilon)^{1/2} \sin \pi (\teta_1-\gamma) - e^{-\tlambda r} (2\tlambda r)^{\teta_1} D(\epsilon)^{-1/2} (-1)^{-\gamma+\teta_1}\right]
    \end{pmatrix} \\[4mm]
    r\psi^o_{2} \to \frac{A^{1/2}(\epsilon)}{\sqrt{2\pi \tlambda}}  
    \begin{pmatrix} \sqrt{\epsilon+1} \left[-e^{\tlambda r} (2\tlambda r)^{-\teta_1} D^*(\epsilon)^{1/2} \sin \pi (\teta_1+\gamma) + e^{-\tlambda r} (2\tlambda r)^{\teta_1} D^*(\epsilon)^{-1/2} (-1)^{\gamma+\teta_1}\right] \\[3mm] 
    \sqrt{1-\epsilon} \left[-e^{\tlambda r} (2\tlambda r)^{-\teta_1} D^*(\epsilon)^{1/2} \sin \pi (\teta_1+\gamma) - e^{-\tlambda r} (2\tlambda r)^{\teta_1} D^*(\epsilon)^{-1/2} (-1)^{\gamma+\teta_1}\right]
    \end{pmatrix}
  \end{aligned},
\end{equation} 

Similarly to the analysis at positive energies, the most convenient choice for the irregular solution is given by the linear combination $\psi^o_3=A(\epsilon) \cot 2\pi\gamma \psi^o_1-\csc 2\pi\gamma \psi^o_2$. This function has the asymptotic form

\begin{equation}
  \psi_3^o \to \frac{A^{1/2}(\epsilon)}{\sqrt{2\pi \tlambda}} \begin{pmatrix}
        \sqrt{\epsilon+1} \left[e^{\tlambda r} (2\tlambda r)^{-\eta_1} D^{1/2}(\epsilon) \cos \pi(\eta_1-\gamma)- e^{-\tlambda r} (2\tlambda r)^{\eta_1} D^{-1/2}(\epsilon) i (-1)^{\teta_1-\gamma} \right] \\[3mm]
        \sqrt{1-\epsilon} \left[e^{\tlambda r} (2\tlambda r)^{-\eta_1} D^{1/2}(\epsilon) \cos \pi(\eta_1-\gamma)+ e^{-\tlambda r} (2\tlambda r)^{\eta_1} D^{-1/2}(\epsilon) i (-1)^{\teta_1-\gamma} \right] 
    \end{pmatrix}
\end{equation}

The functions that extend the energy normalized solutions to negative energies will simply be given by $\frak{f} = A^{1/2} \frak{f}^o$ and $\frak{g} = A^{-1/2} \frak{g}^o$. In the numerical implementation we used the functions that are analytical in energy, and only used the energy normalized functions when calculating quantities related to physical solutions in the open channels. 

With these functions, two quantum defects can be defined, as any solution at negative energies will be expanded as 
\begin{equation}
  \begin{aligned}
    \psi = \frak{f} \cos \pi \mu - \frak{g} \sin \pi \mu = A^{1/2} (\frak{f}^o \cos \pi \mu^o - \frak{g}^o \sin \pi \mu^o )
  \end{aligned} 
\end{equation}
With this formulation, the relation between the two quantum defects is given by 
\begin{equation}
    \tan \pi \mu^o = \frac{\tan \pi \mu}{A}
\end{equation}
Using the asymptotic forms, the bound energies for the single channel Coulomb problem are found when the function $\teta_1-\gamma+\mu$ equals an integer. Equivalently, these energies are found at the points where $A \tan \pi \mu^o + \tan \pi (\teta_1-\gamma) = 0$. These findings can be combined into table \ref{tab:genmqdt} where we present the generalized quantum defect parameters as was done by Greene, et al \cite{Greene1979}.

\begin{table}[]
  \centering
  (A) Parameters
  
  \begin{tabular}{c  c}
  \hline\\[2mm]
       $A(\epsilon,\gamma)$ &  $\frac{\Gamma(\gamma+\teta_1)}{\Gamma(1+\teta_1-\gamma)} \frac{\kappa \epsilon-\gamma}{\kappa-\gamma} \teta_2^{1-2\gamma}$ \\[8mm]
       $B(\epsilon,\gamma)$ &  $|\Gamma(\gamma+i\eta_1)|^2 \frac{\gamma-\kappa \epsilon}{2\pi(\gamma-\kappa)} \teta_2^{1-2\gamma} e^{\pi \eta_1} $\\[8mm]
       $\beta(\epsilon,\gamma)$ & $\pi(\teta_1-\gamma)$ \\[8mm]
       $\zeta$ & $\teta_1$\\[8mm]
       $\mathcal{G}(\epsilon,\gamma), \epsilon>1$ & $-\csc(2\pi\gamma) e^{-2\pi\eta_1}$\\[8mm]
       
  \end{tabular}
  \\[5mm]
  (B) Transformation between base pairs
  
  \begin{tabular}{c}
  \hline\\[2mm]
   $ \begin{pmatrix}
      \frak{f} \\ \frak{g}
  \end{pmatrix} = \begin{pmatrix}
      B^{1/2} & 0 \\
      B^{1/2} \mathcal{G} & B^{-1/2} \\
  \end{pmatrix} \begin{pmatrix}
      \frak{f}^o \\ \frak{g}^o
  \end{pmatrix}, \epsilon>1$ \\[5mm] $ \begin{pmatrix}
    \frak{f} \\ \frak{g}
\end{pmatrix} = \begin{pmatrix}
    A^{1/2} & 0 \\
    0 & A^{-1/2} \\
\end{pmatrix} \begin{pmatrix}
    \frak{f}^o \\ \frak{g}^o
\end{pmatrix}, \epsilon<1$ \\
  \end{tabular}
  \caption{Summary of formulas for the generalized quantum defect parameters}
  \label{tab:genmqdt}
\end{table}

\section{Multichannel Formulation}

If we have multiple channels, the form of the wave function will follow the typical non-relativistic form. As we discuss in the main text, one of the solutions obtained with the variational principle when projected onto the channel function will have the form shown in Eq.~\ref{eq:multichanbeta}, and by using the inverse of the matrix $\mathcal{I}_{m\zeta}$ one defines the solutions in terms of the reaction matrix $K$, for any $r>R$, as
\begin{equation}
  r \ket{\Psi_{j'}} = \mathcal{A} \sum_j \Phi_{j} \left[\frak{f}_{j}^o \delta_{j,j'} - \frak{g}_j^o(r) K_{j,j'}\right]
\end{equation}
where $\mathcal{A}$ is the antisymmetrization operator. We want to use these functions to obtain the eigenchannel solutions that are defined to have the same phase shift on every open channel and have no exponential divergence in the closed channels. Therefore, we need to find the coefficients $Q_{j' \rho}$ that expand these states
\begin{equation}
    \ket{\Psi_\rho} = \sum_{j'} \ket{\Psi_{j'}} Q_{j' \rho} 
\end{equation}
that follow the equations
 \begin{equation}
  \begin{aligned}
    \sum_{j'} \left(-A_{j}^{-1/2} \sin \beta_j \delta_{j j'} - A_{j}^{1/2} \cos \beta_j K_{j j'} \right) Q_{j' \rho} = 0 \quad \text{ if $j$ is closed} \\
    \sum_{j'} \left(\frak{f}^o_j \delta_{j j'} - \frak{g}^o_j K_{j j'} \right) Q_{j' \rho} = T_{j\rho} \left( \frak{f}_{j} \cos \pi \tau_\rho - \frak{g}_{j}\sin \pi \tau_\rho \right) \quad \text{ if $j$ is open}
  \end{aligned}
 \end{equation}
 with $T_{j \rho}$ being a unitary matrix so that the desired boundary conditions are met. The arbitrariness in the normalization of the coefficients $Q$ is fixed by normalizing $T$ to be unitary.\\
 
 One can simplify the equation for the closed channels by multiplying by the matrix with $-A_{j}^{1/2}$ in the diagonal. In the case of the open channels, one must expand the analytic functions in terms of the energy-normalized functions. By separating the terms that go with $\frak{f}$ and $\frak{g}$, and combining the two equations in the usual way to eliminate $T$ one arrives at the generalized eigenvalue problem 
 \begin{equation}
  \begin{aligned}
    \sum_{j'} \left[\sin \beta_j + A_{j} \cos \beta_j K_{j j'}\right] Q_{j'\rho} = 0 \quad \text{if $j$ is closed} \\ 
    \sum_{j'} B_{j} K_{j j'} Q_{j'\rho}  = \tan \pi \tau_\rho \sum_{j'} \left(\delta_{j j'} + B_{j} \mathcal{G}_{j} K_{j j'}\right) Q_{j'\rho} \quad \text{if $j$ is open}
  \end{aligned}
 \end{equation}

 The solution to these equations, and the $T$ matrix, will entirely determine the physical states that are combined identically to the non-relativistic theory to find the physically relevant states. Once the coefficients are found, the normalization matrix can be found by using
 \begin{equation}
     T_{j \rho} = A^{-1/2}_j \sin \pi \tau_\rho Q_{j\rho} + A^{1/2}_j (\sin \pi \tau_\rho + \cos \pi \tau_\rho \mathcal{G}_j) \sum_{j'}K_{j,j'} Q_{j'\rho}
 \end{equation}

To describe photoionization events, the appropriate wave functions have a traveling wave in a single channel going to the detector. These solutions, also known as the incoming wave boundary solutions \cite{Aymar1996}, are expanded in terms of eigenchannel solutions as 
 \begin{equation}
     \ket{\Psi_j^-} = \sum_{\rho} T^\dagger_{\rho j} \ket{\Psi_\rho} e^{-i(\pi \tau_\rho-\sigma_j-\gamma_j \pi/2)} 
 \end{equation}
These solutions correspond to an electron reaching the detector in a specific open channel, but if there are closed channels at that energy we use the $Z$ coefficients to describe the density of states in the closed channel components and thus help in the classification of autoionizing resonances. These coefficients are found by isolating the coefficient that multiplies the exponentially decaying function in the closed channel. For an open channel $j$, we have
 \begin{equation}
 \begin{aligned}
     e^{-i\sigma_j-i\gamma_j\pi/2}\ket{\Psi_j^{-}} &= \sum_{\rho}e^{-i\pi \tau_\rho} T^\dagger_{\rho j} \sum_{j'} \ket{\Psi_{j'}} Q_{j'\rho} \\
     & \underset{r>R}{=}\mathcal{A}\sum_{b} \ket{\Phi_b} \left[\sum_{\rho}e^{-i\pi \tau_\rho} T^\dagger_{\rho j} \sum_{j'}   (\mathfrak{f}^o_b \delta_{b j'}-\mathfrak{g}^o_b K_{b j'}) Q_{j' \rho}\right]
 \end{aligned}
 \end{equation}
where $\sigma_j$ is the Coulomb phase described in the previous appendix. The sum over $\rho$ sums over the eigenchannels, and the sum over $b$ goes over all channels, as does the sum over $j'$.

To explore the asymptotics of this function, we need to separate the sum over $b$ in open and closed channels and replace the asymptotic forms for the two Coulomb functions and the properties of the $Q$ coefficients. Doing so leads us to the following expansion,
\begin{equation}
\begin{aligned}
    \ket{\psi_j^-} \underset{r\to\infty}{\to} \mathcal{A} \left\{ \sum_{b\text{ is open}} \Phi_b \frac{1}{2\sqrt{2 \pi \lambda_b}}\left[e^{i\lambda_b r}  \begin{pmatrix}
        \sqrt{1+\epsilon_b} \\[3mm] i \sqrt{\epsilon_b-1} \end{pmatrix}  \delta_{bj} + \sum_{\rho} e^{-i\lambda_b r} \begin{pmatrix} \sqrt{1+\epsilon_b} \\[3mm] -i \sqrt{\epsilon_b-1} \end{pmatrix} e^{i\sigma_b+i\gamma_b \pi/2}T_{b\rho}e^{-2 i \pi \tau_\rho}T^\dagger_{\rho j}e^{i\sigma_j + i\gamma_j \pi/2} \right]\right. \\[2mm]
        \left. + \sum_{b\text{ is closed}} \Phi_b \frac{1}{\sqrt{2\pi \tlambda_b}} e^{-\tlambda_b r}(2\tlambda_b r)^{\teta_{1b}}D_b^{-1/2} %
        \begin{pmatrix}
        \sqrt{1+\epsilon_b} \\[3mm] -\sqrt{1-\epsilon_b}   
        \end{pmatrix} Z_{b j}
        \right\} 
\end{aligned}
\end{equation}

Explicitly, the $Z$ coefficients are given by
\begin{equation}\label{eq:zcoeffeq}
     Z_{bj} = e^{i\sigma_j+i\gamma_j\pi/2} \sum_{j' \rho} \left(A_b^{-1/2} \cos \beta_b \delta_{b j'} - A_b^{1/2} \sin \beta_b K_{bj'}\right) Q_{j' \rho} e^{-i\pi \tau_\rho} T^\dagger_{\rho j}
\end{equation}

Analyzing where this function peaks, and for which closed channel, as a function of energy, allows for the categorization and labeling of autoionizing resonances. 
\end{widetext}
\end{document}